\documentstyle[prb,aps,epsf]{revtex}
\begin{document}
\begin{center}
{\large\bf  Perturbative studies of the conductivity}\\
{\large\bf in the vortex-liquid regime}\\
\vspace{3ex}
T. Blum$^{*\dagger}$ and M. A. Moore$^*$ 
\\
{\em $^*$Department of Physics, University of Manchester,}\\
{\em Manchester, M13 9PL, United Kingdom.}\\
{\em $^{\dagger}$Department of Physics, Jesse W. Beams 
Laboratory,}\\ 
{\em McCormick Rd., University of Virginia,}\\
{\em Charlottesville, VA 22901.}\\
(\today )
\end{center}
\begin{abstract}
We calculate the Aslamazov-Larkin term of the conductivity in 
the presence of a magnetic field applied along the $c$ axis 
from the time-dependent Ginzburg-Landau equation perturbatively 
using two approaches.  
 In the first a uniform electric field is explicitly applied; 
in the second the Kubo formula is used to extract the linear 
response. 
 The former yields a version of the flux-flow formula for the 
uniform $ab$-plane conductivity, $\sigma_{xx}({\bf k=0})$, that 
holds to all orders of perturbation theory. 
 Obtaining the same result from the Kubo formula requires 
considerable cancellation of terms. 
 We also use the Kubo calculation to examine the nonlocal 
$ab$-plane conductivity, $\sigma_{xx}({\bf k \neq 0})$ (where 
the cancellations no longer occur), as well as the nonlocal 
$c$-axis conductivity $\sigma_{zz}({\bf k \neq 0})$, and look 
for the perturbative precursors of the growing viscous length 
scales. 
 In addition, we consider the effects of weak disorder --- 
both uncorrelated (point defects) and correlated (columnar 
and planar defects). 
\end{abstract}


\section{Introduction}

 Recent experiments on $YBCO$ have exploited nonuniform current 
distributions to reveal the nonlocal nature of the conductivity 
in the vortex-liquid regime.
 The conductivity is nonlocal if the current at a site ${\bf r}$ 
is determined not only by ${\bf E}({\bf r})$ the electric field 
at ${\bf r}$ but also by the fields at other sites ${\bf E}
({\bf r}^{\prime})$. 
 The conductivity is always nonlocal on some microscopic scale, 
but the term ``nonlocal" is generally reserved for cases in 
which the length scale involved is much longer than some 
underlying microscopic one.
 A nonlocal conductivity implies a nonlocal resistivity, though 
the length scales on which each varies may differ. 
 Nonlocality may result if the ``integrity" of the vortex along 
the $c$ axis is maintained ({\it i.e.} not much cutting and 
reconnecting) as then a current applied at the top of a sample 
will yield a response not only at the top but also at the bottom.   
 The interactions between vortices may also lead to nonlocal 
effects within the $ab$ plane. 

The nonlocal nature of the conductivity has been studied by 
Safar {\it et al.} \cite{Safar} 
 They injected small currents into the top of a sample with 
a magnetic field applied along the $z$ axis. 
 They extracted the current in one case from a second contact 
along the top and in another case from the bottom. 
 In each configuration a series of voltage differences were 
taken.  
 The results were then analyzed assuming a local though 
anisotropic conductivity ($\sigma_{xx} \neq \sigma_{zz}$) --- 
the so-called modified Montgomery analysis.  
 Safar {\it et al.} \cite{Safar} found an effective value for 
the ratio $\sigma_{xx}/\sigma_{zz}$ for each configuration and 
discovered a discrepancy of order $\sim 10^5$ between them. 
 The inadequacy of the local analysis is taken as evidence for 
a nonlocal conductivity.  
 Similar evidence for nonlocal effects in BSCCO has been 
reported \cite{Keener}.  

It should be pointed out that the nonlocal interpretation has 
been questioned both on theoretical \cite{Brandt} and 
experimental \cite{Elstev} grounds.  
 It should also be noted that the large effects seen by Safar 
{\it et al.} \cite{Safar} were in heavily twinned samples. 
 Evidence for nonlocal effects in pure samples is somewhat 
weaker. 
 L\'opez {\it et al.} \cite{Lopez} made a direct comparison 
of twinned and untwinned samples.  
 In the first of the current configurations described above 
(current into and out of the top), they took voltage 
measurements across the top and bottom of their samples as 
a function of temperature. 
 In twinned samples of YBCO, the resulting curves $V_{top}$ 
and $V_{bot}$ became almost indistinguishable at temperatures 
above the melting transition, indicative of a long $c$-axis 
length scale in the vortex liquid regime. 
 In untwinned crystals, on the other hand, the curves only 
met at the melting transition. 
 This result demonstrates the enhancement of the vortex-line 
integrity and hence nonlocality by the correlated defect.  

On the theoretical side nonlocal conductivity has been studied 
within the hydrodynamic framework\cite{MarNel}.  
 We are assuming a linear but nonlocal relationship between 
current and electric field, {\it i.e.} an integral version 
of Ohm's law
\begin{equation}
j_{\mu}({\bf r}) = \int \sigma_{\mu \nu} ({\bf r}, 
{\bf r^{\prime}}) E_{\nu}({\bf r^{\prime}}) 
d {\bf r^{\prime}}, 
\label{Ohm}
\end{equation}
where 
$\sigma_{\mu \nu}({\bf r},{\bf r^{\prime}}) \neq 
\sigma_{\mu \nu} \delta ({\bf r} - {\bf r^{\prime}})$.
If the system is translationally invariant, Fourier 
transforming equation (\ref{Ohm}) results in 
\begin{equation}
j_{\mu}({\bf k}) =  \sigma_{\mu \nu} ({\bf k})~E_{\nu}({\bf k}), 
\label{Ohm-2}
\end{equation}
where nonlocality now implies $\sigma_{\mu \nu}({\bf k}) 
\neq {\rm const.}$ 
 The emphasis is on the large-distance behavior in real space 
and so presumably on the small-${\bf k}$ behavior of the 
Fourier transform. 
 Huse and Majumdar\cite{HusMaj} explored a hydrodynamic approach, 
in which the small-${\bf k}$ expansion of $\sigma_{\mu \nu}
({\bf k})$ is truncated as follows: 
\begin{equation}
\sigma_{\mu \nu}({\bf k}) \ = \ \sigma_{\mu \nu}(\bf{0})
\ + \ S_{\mu \alpha \beta \nu}~k_{\alpha}k_{\beta}. 
\label{hydro}
\end{equation}
In the example they work out in detail to explain features seen 
by Safar {\it et al.}, they include only one $S$: $S_{xzzx}$. 
 It models the ``tilt viscosity" --- which measures the influence 
of pancake vortices moving in one $ab$ plane on those moving in 
another when they travel with different velocities.\cite{HusMaj}

Stability requires that $\sigma_{\mu \nu}({\bf k})$ be a positive 
definite matrix.  
 For the hydrodynamic form of $\sigma_{\mu \nu}({\bf k})$, it 
implies, for example, that $S_{xzzx}$ and $S_{xxxx}$ must be 
positive. 
 When these conditions are met, one finds that the 
{\it resistivity} decays with a length scale of order 
($\sqrt{S/\sigma}$), which is in turn the length scale of the 
electric-field and current distributions.\cite{HusMaj} 
 (The hydrodynamic approach also predicts surface currents which 
supposedly should be spread out over some small length scale not 
accessible by that technique.) 
 Recent calculations\cite{Mouetal,BM} starting from the 
time-dependent Ginzburg-Landau (TDGL) equation find that many of 
the $S$'s do not have the requisite sign --- at least not for 
clean samples at high temperatures. 
 Those calculations were taken only to Gaussian order (lowest 
order in perturbation theory); the purpose of the present paper 
is to go beyond Gaussian order and also to explore the effect 
of various sorts of defects: point, columnar and planar. 
 Recent simulations of the TDGL equations in two dimensions did 
find a region above the melting transition in which $S_{xxxx} 
\geq 0$.  \cite{Wortis}  
 (The problem of calculating the voltage distribution whether 
or not the conductivities are of the ``hydrodynamic" form will 
be presented elsewhere. \cite{Phillipson})

The first order of business is to calculate the uniform 
conductivity $\sigma_{\mu \nu}(\bf{0})$.
 The resistivity of clean samples of YBCO in the vortex-liquid 
regime varies smoothly with temperature until it experiences a 
sudden drop generally associated with the first-order melting 
transition to a pinned vortex solid. 
 The temperature and field dependences in this smooth region 
agree with the flux-flow formula \cite{Kim} which was originally 
conceived of in the vortex-solid phase where one pictures the 
translation of the entire unpinned vortex lattice. 
 So how is it that the same physics applies in the vortex-liquid 
regime?
 The important feature of the flux-flow formula besides the 
absence of pinning by defects would seem to be its insensitivity 
to viscous effects. 
 As it is the entire vortex system that moves, the amount of 
entanglement along the $c$-axis or the extent of crystalline order 
in the $ab$ plane is irrelevant. 
 It takes a nonuniform current distribution to probe such viscous 
effects.   
 
 We will determine $\sigma_{\mu \nu}(\bf{0})$ in two ways.  
 First, we will consider the TDGL equation with an explicit 
uniform electric field applied, calculating the current and 
extracting $\sigma_{\mu \nu}(\bf{0})$ directly. 
 Second, we will calculate the conductivity from the Kubo formula, 
a version of the fluctuation-dissipation theorem which yields the 
linear response of a system not too far from equilibrium. 
 The electric field does not explicitly appear in the latter.  
 We will see that a delicate cancellation of terms arising in the 
Kubo calculation is necessary for the two approaches to agree, and 
we will demonstrate this cancellation to second order in 
perturbation theory for the conductivity of two-dimensional 
samples.  
 We will then proceed to examine the nonlocal effects within the 
Kubo formalism, see how the ${\bf k=0}$ cancellation breaks down 
at ${\bf k \neq 0}$ and show how viscous effects may enter.

Next we consider the effect of disorder --- first on the uniform 
conductivity. 
 It would seem to have two competing effects. 
 On one hand, defects pin the vortices and thus might lower the 
resistivity. 
 On the other hand, they disrupt the formation of a vortex lattice 
\cite{Larkin} which is more readily pinned than individual 
vortices, and thus they might raise the resistivity. 
 In fact, both effects are seen in the experiments of Fendrich 
{\it et al.} \cite{Fendrich} in which point defects were induced 
by electron irradiation. 
 Introducing the defects eliminated the sudden drop in resistivity 
at $T_{c}$. 
 At temperatures above $T_{c}$, the irradiated sample had a lower 
resistivity; while at temperatures below $T_{c}$, it had a higher 
resistivity.  
 Another outcome of inducing the defects is that the temperature 
and field dependence is no longer of the flux-flow type. 
 At low temperatures an activated form might be expected 
\cite{activated-form}, but such considerations are beyond the 
scope of the present work which employs standard perturbation 
theory.  
 We will examine the effect of these defects in the weak disorder 
limit.  

 Even the lowest-order calculation of $\sigma_{\mu \nu} ({\bf k}, 
\omega)$, the wave-vector and frequency-dependent conductivity, 
is cumbersome. \cite{Mouetal,BM}   
 To simplify the results we will concentrate on the leading 
behavior.
 In the presence of a magnetic field, Landau levels provide a 
natural basis for calculating and describing various phenomena. 
 Two energy scales naturally arise: the first is $\alpha_H$, 
the energy of states in the $n=0$ or lowest Landau level (LLL); 
the second is the energy spacing between Landau levels given 
by $\hbar \omega_0$ with $\omega_0$ the cyclotron frequency. 
 We will gain an immense simplification by focusing on the 
regime in which $\alpha_H  \ll \hbar \omega_0$, the so-called 
LLL approximation. 
 Arguments based on the {\it renormalized} values of $\alpha_H$ 
and $\hbar \omega_0$ suggest that such a separation of scales 
holds over a significant portion of the vortex-liquid regime. 
\cite{Ikeda1}
 Note that $\alpha_H  \ll \hbar \omega_0$ limit does not imply 
that only $n=0$ states are employed in a calculation, although 
it usually does imply that the number of $n \neq 0$ states is 
kept to a minimum. 
 Considering only fluctuations in the LLL states serves not 
only to simplify the calculations but also to regularize them. 

Before presenting the calculations in detail let us highlight 
a couple key features.  
 The Kubo formula for the conductivity involves the product 
of two current densities, and each current density involves 
a product of a $\Psi$ and a $\Psi^*$, where $\Psi$ is the 
superconducting order parameter. 
 (See eqs. (\ref{kubo1}) and (\ref{kubo2}) below.) 
 Thus, calculated in this way the conductivity is a 
``four-point" object, that is, it involves four $\Psi$ fields. 
 If these $\Psi$'s are expanded in the Landau-level basis, a 
crucial question arises: how many of these four $\Psi$'s can 
be in the LLL? 
 Already we find a notable distinction between the $c$-axis 
conductivity, $\sigma_{zz}({\bf 0})$, for which the answer is 
all four and the $ab$-plane conductivity, $\sigma_{xx}
({\bf 0})$, for which the answer is only two. 
 As a result, in the LLL regime the characteristic time scale 
inherent in $\sigma_{zz}$ is much larger than that in 
$\sigma_{xx}$. 
 Another outcome is that in terms of LLL states ($\sigma_{xx}
({\bf 0})$) is a two-point quantity and as such is independent 
of viscous effects which are related to four-point quantities. 
 
 Understanding the effect of random pinning in the 
vortex-liquid regime is no trivial matter. \cite{Blatter}
 When point disorder is added, translational invariance is 
destroyed. 
 Moreover, while the division of $\sigma_{xx}({\bf 0})$ into 
$n=0$ and $n \geq 1$ parts persists, the $n=0$ component now 
becomes a four-point object sensitive to viscous effects.  
 As a result the associated time and length scales may grow. 
 The separation of $n=0$ and $n \geq 1$ states also plays a 
role in the distinction between $c$-axis length scales in 
materials with correlated and uncorrelated disorder. 
 To each energy level there is associated a $c$-axis length 
scale, and the one linked  to the $n=0$ level is much longer. 
 It turns out that the $c$-axis length scale of $\sigma_{xx}$ 
is controlled by the $n=0$ states in the presence of columnar 
and planar defects but by the $n \geq 1$ states in the presence 
of point defects. 

The rest of the paper is organized as follows.
 In the next section we lay the groundwork for the calculations 
that follow by constructing the Green's function and 
correlation function from the TDGL equation. 
 In the section following that we calculate the uniform 
conductivity, $\sigma_{xx} ({\bf 0})$, first by explicitly 
applying the electric field and then via the Kubo formula. 
 We then move on to consider the nonlocal conductivity 
$\sigma_{xx}({\bf k})$. 
 After that we examine the conductivity of films.  
 We then consider the effect of weak disorder and finally 
summarize.

\section{Time-Dependent Ginzburg-Landau Theory}

We begin with the usual Ginzburg-Landau free-energy 
functional ${\cal F}[\Psi]$  for the superconducting order 
parameter $\Psi({\bf r},t)$ in a magnetic field
\begin{equation}
{\cal F}[\Psi]= \int d^3{\bf r} \Biggl[\sum_{j=x,y,z} 
{  \left|P_{j} \Psi \right|^2 \over 2 m_j} 
+\alpha |\Psi|^2 + {\beta \over 2} |\Psi|^4 
+{1 \over 2 \mu_0}({\bf \nabla} \times {\bf A})^2\Biggr].
\label{free-energy}
\end{equation}
where ${\bf A}$ is the vector potential and $P_{j}$ is the 
$j^{th}$ component of the momentum operator given by 
\begin{equation}
P_{j} = {\hbar \over i} {\partial \over \partial r_{j} } 
-e^* A_{j}({\bf r})
\label{momentum}
\end{equation}
with $e^*=2e$. 
 This combination arises for reasons of gauge invariance. 
 We allow for two masses $m_{x,y}=m_{ab}$ and $m_z=m_c$ to 
reflect to some extent the anisotropy of  high $T_c$ 
materials; of course, the calculations can be extended 
to the Lawrence-Doniach model in which the layering is 
treated more explicitly. 
 
 Next, we choose a simple relaxational dynamics for $\Psi$ 
given by the time-dependent Ginzburg-Landau equation
\begin{equation}
{ 1 \over \Gamma}
\left( {\partial  \over \partial t} +{ie^* \Phi({\bf r},t) 
\over \hbar} \right)\Psi ({\bf r},t)= - {\delta {\cal F}
[\Psi] \over \delta \Psi^*({\bf r},t) }+ \eta({\bf r},t).  
\label{TDGL}
\end{equation}
 The field $\Phi({\bf r},t)$ is related to the chemical 
potential which accompanies the time derivative in order 
to maintain gauge invariance; in what follows we will use 
the approximation that $\Phi({\bf r},t)$ is the scalar 
electric potential. \cite{Schmid,Gorkov,Dorsey}
 The thermal fluctuations are represented by the noise 
$\eta({\bf r},t)$, which has zero average and 
$\delta$-function correlations 
\begin{equation}
\left\langle \eta^*({\bf r},t)~\eta({\bf r^{\prime}},
t^{\prime})\right\rangle = {2 k_BT \over \Gamma} 
~\delta({\bf r}-{\bf r^{\prime}}) ~\delta(t - t^{\prime}).
\label{noise} 
\end{equation}
The noise strength is chosen so that in the absence of a 
driving electric field, the distribution of $\Psi$'s, 
${\cal P}[\Psi]$, evolves toward its  equilibrium solution  
\begin{equation}
{\cal P}[\Psi] \propto {\rm exp} \left\{ 
- {1 \over k_B T} \int d^3{\bf r} ~{\cal F}[\Psi] 
\right\}.   
\label{dist}
\end{equation}
We take $\Gamma$, the kinetic coefficient, to be real.
 Note that a complex kinetic coefficient ($\Gamma^{-1} 
\rightarrow \Gamma_0^{-1} + i \lambda_0^{-1}$ in eq. 
(\ref{TDGL}) but $\Gamma^{-1} \rightarrow \Gamma_0^{-1}$ 
in eq. (\ref{noise})) is required to model the Hall 
conductivity $\sigma_{xy}$. 
\cite{Dorsey,Ullah}

The TDGL equation is supplemented by 
\begin{equation}
\nabla \times \nabla \times {\bf A} = \mu_0 
\left[{\bf J^{(n)}}  + {\bf J^{(s)}} \right], 
\label{Maxwell}
\end{equation}
where ${\bf J^{(n)}}$ is the normal current given by
\begin{equation}
{\bf J^{(n)}} = \sigma^{(n)}  \left[ -\nabla \Phi - 
{\partial {\bf A }\over \partial t} \right]
\label{normal-current}
\end{equation}
and ${\bf J^{(s)}}$ is the superconducting current given 
by 
\begin{equation}
 J_{j}^{(s)}({\bf r},t) = {e^* \over 2 m_{j}}
\left[ P_{1j} +  P_{2j}^* \right] \Psi^*({\bf r_2},t) 
\Psi({\bf r_1},t) \Biggl|_{{\bf r_1}={\bf r_2}={\bf r}}, 
\label{current}
\end{equation}
with $P_j$ is the momentum operator (\ref{momentum}).  
 Eq. (\ref{Maxwell}) is simply Maxwell's equation 
{\it sans} the Maxwell displacement current which is 
presumed negligibly small.   
 Actually in this paper we concentrate solely on the 
TDGL equation, so that certain effects that enter with 
the Maxwell equations, such as backflow 
\cite{Hu+Thompson,Pelcovits,Troy}, will be missing.   

 Let us consider an electric field applied along the $x$ 
direction: $\Phi({\bf r},t)=-E~x$ and a magnetic field 
applied in the $z$ direction: $({\bf A}=Bx{\hat y})$.  
 The TDGL equation then becomes 
\begin{equation}
\left[{1 \over \Gamma} {\partial \over \partial t} 
+{\cal H}({\bf r})\right]\psi({\bf r},t) 
= \eta({\bf r},t)-\beta|\psi ({\bf r},t)|^2 
\psi({\bf r},t) ,
\label{TDGL2}
\end{equation}
where ${\cal H}$ is 
\begin{equation}
{\cal H} =  
-{\hbar^2 \over 2 m_{ab}} \left( {\partial^2 \over 
\partial x^2} +\left({\partial \over \partial y}-
{i e^* B x \over \hbar}\right)^2 \right) - 
{\hbar^2 \over 2m_{c}} {\partial^2 \over \partial z^2}
 +\alpha - {i e^* E x \over \Gamma \hbar}.
\label{hamiltonian}
\end{equation}
For the calculations that follow we will need the Green's 
function which satisfies:
\begin{equation} 
\left[{1 \over \Gamma} {\partial  \over \partial t} 
+{\cal H}({\bf r})\right] 
~G({\bf r},t;{\bf r^{\prime}}, t^{\prime}) 
\ = \ \delta({\bf r}-{\bf r^{\prime}}) ~\delta
(t-t^{\prime}). 
\label{green1}
\end{equation}
The Green's function serves as the inverse of the operator 
$(\Gamma^{-1} \partial_t + {\cal H})$, allowing us to rewrite 
the TDGL equation as
\begin{equation}
\Psi({\bf r},t) = \int d {\bf r^{\prime}} \int d t^{\prime} 
~G({\bf r},t;{\bf r^{\prime}}, t^{\prime}) \left[ 
\eta({\bf r^{\prime}},t^{\prime}) -\beta |\Psi
({\bf r^{\prime}}, t^{\prime})|^2\Psi({\bf r^{\prime}}, 
t^{\prime}) \right],
\label{TDGL-inverted-2}
\end{equation} 
which we can write in a more symbolic form 
\begin{equation}
\Psi_1 = G_{1,2} \eta_2 -\beta G_{1,2}\Psi^*_2 \Psi_2 \Psi_2.
\label{symbol}
\end{equation}
and solve --- at least formally --- by iteration 
\begin{equation}
\Psi_1 = G_{1,2} \eta_2 
-\beta G_{1,2}G_{2,3}^*G_{2,4}G_{2,5}\eta_3^* \eta_4 \eta_5 
+O(\beta^2).
\label{iterate-b}
\end{equation}
This expansion forms the basis for the standard perturbation 
theory in $\beta$. 

We can construct $G({\bf r},t;{\bf r^{\prime}}, t^{\prime})$ 
from the eigenstates $\phi_n(k_y,k_z;{\bf r})$ and 
eigenvalues $E_n(k_y,k_z)$ of ${\cal H}$, which are
\begin{equation}
\phi_n(k_y,k_z;{\bf r})= e^{ik_yy+ik_zz} ~u_n\left(
{x \over \ell} - k_y \ell-{ivm_{ab} \ell \over \hbar^2 
\Gamma}\right),
\label{states}
\end{equation}
and
\begin{equation}
E_n(k_y,k_z)={\hbar^2 k_z^2 \over 2m_c} + 
\underbrace{\alpha + {\hbar \omega_0 \over 2}+
{m_{ab}v^2 \over 2 \hbar^2 \Gamma^2}}_{\alpha_{EH}} 
+n~ \hbar \omega_0 -{i k_y v \over \Gamma}, 
\label{energy}
\end{equation}
where  
\begin{equation}
\ell = \left({\hbar \over e^*B}\right)^{1/2}, \ \ \  
\omega_0={e^*B \over m_{ab}} \ \ \ {\rm and} \ \ \ 
v={E \over B}.
\end{equation}
($\ell$ is the magnetic length and is associated with the 
distance between vortices; $\omega_0$ is the cyclotron 
frequency; and $v$ is roughly speaking the speed at which 
the flux lines move.)   
The functions $u_n(s)$ are given by 
\begin{equation}
u_n(s) = {H_n(s)~ {\rm exp}\{-s^2/2 \} \over 
\left(\ell \sqrt{\pi} 2^n n! \right)^{1/2} },  
\label{harmonic}
\end{equation}
where $H_n(s)$ are the Hermite polynomials. 

Note that in the absence of an electric field ($v=0$) the 
energy does not depend on $k_y$ giving the characteristic 
large degeneracy of the Landau levels. 
 Moreover, we then find in expression (\ref{energy}) the two 
energy scales mentioned in the introduction: 1)  $\alpha_{H}=
\alpha + \hbar \omega_0/2$, the energy in the LLL at $k_z=0$ 
and 2) $\hbar \omega_0$  the energy spacing between levels. 
 Since the temperature at which $\alpha_H=0$ is where one 
expects the LLL modes to go critical within mean-field theory, 
it is taken to define the mean-field $H_{c2}(T)$ line.   

 Given the presence of the $k_z^2$ term in the energy, it 
is convenient to construct two $c$-axis length scales, one 
corresponding to each energy scale; they are 
\begin{equation}
\xi_c = \left({\hbar^2 \over 2 m_c \alpha_H}\right)^{1/2} 
\ \ \ \ 
{\rm and} 
\ \ \ \ 
\ell_c = \left( { \hbar \over m_c \omega_0} \right)^{1/2}. 
\label{c-axis-lengths}
\end{equation}  
The former is the standard mean-field $c$-axis correlation 
length, which is temperature dependent; while the latter is 
a $c$-axis version of the magnetic length 
\begin{equation}
\ell_c = \left( {m_{ab} \over m_c} \right)^{1/2} \ell, 
\label{c-lengths}
\end{equation}  
which is temperature independent. 

From the eigenstates and eigenvalues above we construct the 
following Green's function 
\begin{eqnarray}
G({\bf r},t;{\bf r^{\prime}}, t^{\prime}) = &&
\Gamma \int{d \omega \over 2 \pi}
\int {d k_y \over 2 \pi} \int {d k_z \over 2 \pi}  
~{\rm exp}\left\{ik_y(y-y^{\prime}) +ik_z(z - z^{\prime})
-i\omega(t-t^{\prime}) \right\} \nonumber \\
&& \times \sum_{n=0}^{\infty} ~ {u_n\left({x \over \ell} 
-k_y \ell -i \tilde k_v \ell \right)
u_n\left({x^{\prime} \over \ell} -k_y \ell -i \tilde k_v 
\ell \right)\over \Gamma E_n(k_y,k_z) -i \omega }, 
\label{green3-1}
\end{eqnarray}
where $\tilde k_v= m_{ab} v / \hbar^2 \Gamma$. 
 One might notice that when $v \neq 0$ the arguments of 
the $u$'s are not complex conjugates with $x 
\leftrightarrow x^{\prime}$; that is because the operator 
${\cal H}$ is not Hermitian.  
 The real issue is that eq. (\ref{green1}) is satisfied.   

Various representations of $G({\bf r},t;{\bf r^{\prime}}, 
t^{\prime})$ are useful depending on the calculation in 
question.  
 Another useful form of $G({\bf r},t;{\bf r^{\prime}}, 
t^{\prime})$  can be derived by performing the integral 
over $\omega$ and using the identity 
\begin{equation}
\sum_{n=0}^{\infty} {H_n(x)H_n(y) \over 2^n n!}t^n = 
{{\rm e}^{[2xyt-(x^2+y^2)t^2]/(1-t^2)} \over 
(1-t^2)^{1/2} }.
\label{identity}
\end{equation}
It leads to   
\begin{eqnarray}
G({\bf r},t;{\bf r^{\prime}}, t^{\prime}) = 
&&{\Gamma \over 4 \pi \ell^2 } \int {dk_z \over 2 \pi} 
\left[{\rm sinh}\left[{\Gamma \hbar \omega_0 (t-t^{\prime}) 
\over 2}\right] \right]^{-1}
{\rm exp}\left\{-\Gamma\left[{\hbar^2k_z^2 \over 2 m_c} 
+{\hbar^2 {\tilde k_v}^2 \over 2 m_{ab}}+ \alpha 
\right](t-t^{\prime}) \right\}
\nonumber \\
&& {\rm exp}\left\{- 
{\rm coth}\left[{\Gamma \hbar \omega_0 (t-t^{\prime}) 
\over 2}\right] 
{\left[(x-x^{\prime})^2 +(y-y^{\prime}+
v(t-t^{\prime}))^2 \right] 
\over 4 \ell^2} \right\}
\nonumber \\
&&
{\rm exp}\left\{{i(x+x^{\prime}-2i \tilde k_v\ell^2 )
(y-y^{\prime}+v(t-t^{\prime})) \over 
2 \ell^2}+ ik_z(z-z^{\prime})\right\}\Theta(t-t^{\prime}) .
\label{green4}  
\end{eqnarray}
 In this form there is no longer any summation and the spatial 
dependence is simply that of a Gaussian, but the price paid is 
in the hyperbolic time dependence.   
 One can see in the combination $(y-y^{\prime}+v(t-t^{\prime}))$ 
the tendency for the vortices to move in the negative $y$ 
direction under the influence of the Lorentz force.

After the Green's function, the next quantity we will need is 
the correlation function $C({\bf r},t;{\bf r^{\prime}},
t^{\prime})= \left\langle \Psi({\bf r}, t) ~ \Psi^*({\bf 
r^{\prime}},t^{\prime}) \right\rangle$. 
 There are two complications here:  the operator ${\cal H}$ 
is neither Hermitian nor translationally invariant.   
 As a result some of the formulas relating $C$ and $G$ that 
we have become accustomed to are inappropriate; it is best to 
resort to the definitions.  
 Substituting the expressions for $\Psi({\bf r},t)$ and 
$\Psi^*({\bf r^{\prime}},t^{\prime})$ given by eq. 
(\ref{TDGL-inverted-2}) into the definition of 
$C({\bf r},t;{\bf r^{\prime}},t^{\prime})$ and performing 
the noise average yields  
\begin{equation} 
C({\bf r},t;{\bf r^{\prime}},t^{\prime})  = {2 k_B T  
\over \Gamma} \int dt^{\prime \prime} \int d r^{\prime 
\prime}  ~ G({\bf r},t;{\bf r^{\prime\prime}},t^{\prime 
\prime})G^*({\bf r^{\prime}},t^{\prime};{\bf r^{\prime
\prime}},t^{\prime \prime})  +O(\beta).
\label{correlator}
\end{equation}  
Many of the calculations that follow will use the Kubo 
formula in which there is no explicit electric field ($v=0$). 
 Then the expression for the correlator simplifies somewhat, 
becoming 
\begin{eqnarray}
C_{v=0}({\bf r},t;{\bf r^{\prime}}, t^{\prime}) = &&
2k_B T \Gamma \int {d \omega \over 2 \pi} \int {d k_y 
\over 2 \pi} 
\int {d k_z \over 2 \pi}  ~{\rm exp}\left\{ik_y(y-
y^{\prime}) +ik_z(z - z^{\prime})
-i \omega (t-t^{\prime}) \right\} \nonumber \\
&& \times  \sum_{n=0}^{\infty} 
{u_n\left({x \over \ell} -k_y \ell  \right)
u_n\left({x^{\prime} \over \ell} -k_y \ell \right) 
\over \Gamma^2E_n^2(k_z)+ \omega^2} . 
\label{correlator-2}
\end{eqnarray}
Now let us move on to using these expressions to calculate 
the conductivity.

\section{The fluctuation conductivity} 
 
Having the Green's function and correlation function, we are 
now ready to calculate the conductivity. 
 We will focus on the conductivity due to fluctuations in the 
superconducting order parameter --- the so-called 
Aslamazov-Larkin term. 
 The normal contribution $\sigma^{(n)}$ must be included 
separately, and we  are neglecting other possible contributions 
such as the Maki-Thompson or density-of-states terms.  
\cite{otherterms}

 Recall that conductivities take on a form 
\begin{equation}
\sigma \propto {e^{*2} ~{\cal N} ~\tau \over m^*},
\end{equation}
where $e^*$ and $m^*$ are the effective charge and mass 
respectively, $\tau$ is a characteristic time scale, and 
${\cal N}$ is the carrier density.    
 Since we are interested in the conductivity due to 
superconducting fluctuations $\langle |\Psi|^2 \rangle$ will 
serve as the carrier density.  
 An examination of the TDGL equation reveals that the 
kinetic coefficient $\Gamma$ has dimensions $(energy \times 
time)^{-1}$; so that $time = (\Gamma \times energy)^{-1}$.  
 The question then becomes: what is the appropriate energy 
scale?  
 Thus far two candidates --- $\alpha_H$ and $\hbar \omega_0$ 
--- have emerged. 

\subsection{The direct approach}

 Now let us derive the flux-flow form of the uniform conductivity 
from a calculation with an explicit electric field applied.  
 This approach has several advantages over the Kubo calculation.   
 It extracts the conductivity from the current, a two-point object, 
instead of the current-current correlation function, a 
four-point object. 
 Related to this point is the fact that many of the diagrams 
occurring in the Kubo calculation make canceling contributions.  
 In addition, in the direct approach the $\alpha /\hbar \omega_0 
\rightarrow 0$ limit (the LLL approximation) requires only LLL 
states, which is not true for the Kubo case.  
 It can also be extended to include the nonlinear effects.  
 (The advantage of the Kubo formalism is that it can be used to 
calculate the nonlocal conductivities $\sigma_{xx}({\bf k})$ and 
$\sigma_{zz}({\bf k})$.) 

 From equation (\ref{current}) we see that the average current 
is obtained from the correlation function as follows
\begin{equation}
\langle J_x^{(s)}({\bf r},t)\rangle = {\hbar e^* \over 2 i m_{ab} } 
\left( {\partial \over \partial x} - {\partial \over \partial 
x^{\prime}}\right) C({\bf r},t;{\bf r^{\prime}},t^{\prime}) 
\Biggl|_{({\bf r},t)=({\bf r^{\prime}},t^{\prime})}. 
\label{direct}
\end{equation}
We will drop the superscript $(s)$ hereafter. 
At the lowest order in perturbation in theory, we can insert 
expression (\ref{correlator}) into (\ref{direct}) to obtain 
\begin{equation} 
\langle J_x\rangle   = {k_B T \hbar e^* \over i\Gamma m_{ab}} 
\left( {\partial \over \partial x} - {\partial \over \partial 
x^{\prime}}\right)\int dt^{\prime \prime} \int d r^{\prime 
\prime}  ~ G({\bf r},t;{\bf r^{\prime\prime}},t^{\prime 
\prime})
~G^*({\bf r^{\prime}},t^{\prime};{\bf r^{\prime\prime}},
t^{\prime \prime})
\Biggl|_{({\bf r},t)=({\bf r^{\prime}},t^{\prime})}.   
\label{direct-b0}
\end{equation} 
We have verified a posteriori that in the direct approach the 
$\alpha / \hbar \omega_0 \rightarrow 0$ limit is equivalent to 
using only $n=0$ states from the start.  
 Restricted to $n=0$ states, the Green's function becomes
\begin{eqnarray}
G_0({\bf r},t;{\bf r^{\prime}}, t^{\prime}) = &&
{\Gamma  \over \sqrt{\pi}\ell}
\int {d k_y \over 2 \pi} 
\int {d k_z \over 2 \pi}  
~{\rm exp}\left\{ik_y(y-y^{\prime}) +ik_z(z - z^{\prime})
-\Gamma E_0(k_y,k_z)(t-t^{\prime}) \right\} \nonumber \\
&& \times {\rm exp} \left\{  -{\left({x \over \ell} -k_y 
\ell -i \tilde k_v \ell \right)^2 \over 2} - {\left(
{x^{\prime} \over \ell} -k_y \ell -i \tilde k_v \ell 
\right)^2 \over 2}  \right\}~\Theta(t-t^{\prime}).
\label{green-LLL}  
\end{eqnarray}   
Notice that the derivative with respect to $x$ in eq. 
(\ref{direct-b0}) pulls down a factor of $-(x/\ell^2-k_{y1} 
-i \tilde k_v )$, while that with respect to $x^{\prime}$ 
pulls down $-(x^{\prime}/\ell^2-k_{y2} +i \tilde k_v)$. 
 The wave vectors $k_{y1}$ and $k_{y2}$ will eventually prove 
to be equal (momentum conservation), so when ${\bf r}$ is set 
equal to ${\bf r^{\prime}}$ we find 
\begin{equation}
\langle J_x\rangle ~=~ {\hbar e^* \tilde k_v\over  m_{ab} } 
 C_0({\bf r},t;{\bf r},t) ~=~{ e^{*2}E \over  m_{ab} \Gamma 
\hbar \omega_0 } 
 \langle |\Psi_0|^2 \rangle , 
\label{flux-flow}
\end{equation}
where  $C_0({\bf r},t;{\bf r},t) =\langle |\Psi_0|^2 \rangle$ 
denotes the order-parameter fluctuations in the LLL. 
 (Strictly speaking it is $\langle |\Psi_0|^2 \rangle$ in the 
presence of the electric field, which is where the nonlinear 
effects would enter). 
 Note that the conductivity has the expected form with the 
characteristic time given by $\tau = 1/\Gamma \hbar \omega_0$ and 
that any temperature dependence enters only through the carrier 
density and not through the characteristic time.  

 So far this result is for the lowest order in perturbation 
theory; however, all terms in the perturbative expansion for 
$C({\bf r},t;{\bf r^{\prime}},t^{\prime})$ begin with some 
$G({\bf r},t;{\bf r_1},t_1)$ and end with some 
$G^*({\bf r^{\prime}},t^{\prime};{\bf r_n},t_n)$. 
 It follows that the effect of taking the derivatives in eq. 
(\ref{direct}) will always be the same as it was above, and 
consequently, the flux-flow result, eq. (\ref{flux-flow}), 
holds to all orders of perturbation theory in the LLL 
approximation. 
 This result is remarkable in the simplicity of the relation 
between $\sigma_{xx}({\bf 0})$, a dynamic quantity, and $\langle 
|\Psi_0|^2 \rangle$, a static quantity.  
 One always expects the conductivity to be proportional to carrier 
density, but in this case the proportionality constant $\tau$ turns 
out to be rather trivial --- having no temperature dependence.  

\subsection{The Kubo formalism}

 We now turn to the Kubo approach. 
 For a system not too far from equilibrium the conductivity can 
be calculated from the Kubo formula 
\begin{equation} 
\sigma_{\mu \nu}({\bf k},\omega) \ = \ {1 \over 2 k_B T}
\int d({\bf r}-{\bf r^{\prime}}) \int d(t-t^{\prime}) 
~{\rm e}^{i{\rm k}\cdot ({\rm r}-{\rm r^{\prime}}) 
-i \omega (t - t^{\prime}) }
~\langle J^{(s)}_{\mu}({\bf r},t)~J^{(s)}_{\nu}({\bf r^{\prime}}, 
t^{\prime}) \rangle_c~, 
\label{kubo1}
\end{equation}
where the suffix $c$ denotes the ``connected" piece and where 
this expression allows for a frequency dependence. 
 The Kubo formula relates the conductivity, a dissipative quantity, 
to fluctuations in an associated quantity, here the superconducting 
current. 
 As such, it is a version of a fluctuation-dissipation theorem.   
 It calculates the linear response to an electric field without 
explicitly applying one; the fluctuations above are those in the 
absence of an electric field.  

 Inserting the expression for the current (\ref{current}) into the 
Kubo formula (\ref{kubo1}) yields
\begin{eqnarray}
\sigma_{\mu \nu}({\bf k},\omega) &\ = \ &{e^{*2} \over 8 k_B 
T m_{\mu}m_{\nu}} \int d({\bf r}-{\bf r^{\prime}}) 
\int d(t-t^{\prime}) ~{\rm e}^{i{\rm k}\cdot 
({\rm r}-{\rm r^{\prime}}) -i\omega(t-t^{\prime})} \nonumber \\ 
&&\ 
\Bigl(P_{1\mu}+P_{2\mu}^*\Bigr)\Bigl(P_{3\nu}+P_{4\nu}^*\Bigr)
~\langle  \Psi({\bf r_1},t) \Psi^*({\bf r_2},t) 
\Psi({\bf r_3},t^{\prime})\Psi^*({\bf r_4},t^{\prime})
\rangle_c \Biggl|_{{\bf r_1=r_2=r} \atop 
{\bf r_3=r_4=r^{\prime}}} .
\label{kubo2}
\end{eqnarray}
One can see that calculated in this way the conductivity is a 
four-point object as mentioned in the introduction.  
 Henceforth, we will consider DC results ($\omega=0$) only and 
will drop the frequency dependence from our expressions. 

\begin{figure}
\centerline{
\epsfxsize=7cm \leavevmode \epsfbox{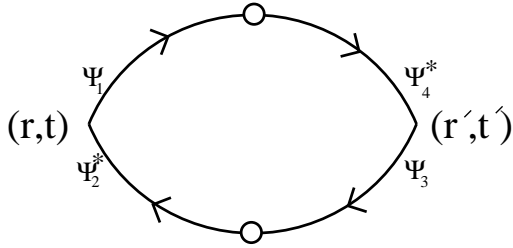}}
\caption{ Gaussian order diagram for the conductivity. 
 An arrow corresponds to a Green's function, and a circle 
corresponds to the noise average.  
 Not drawn but of crucial importance are the momentum operators 
acting at sites $({\bf r},t)$ and $({\bf r^{\prime}},
t^{\prime})$.  }
\label{Gaussian}
\end{figure}

The aim now is to calculate averages of the sort $\langle 
\Psi({\bf r_1},t_1)\Psi^*({\bf r_2},t_2)\Psi({\bf r_3},t_3)
\Psi^*({\bf r_4},t_4) \rangle_c$. 
 At the lowest order in perturbation theory (the Gaussian term); 
one simply applies Wick's theorem  
\begin{equation}
\langle \Psi({\bf r_1},t)\Psi^*({\bf r_2},t)
\Psi({\bf r_3},t^{\prime})\Psi^*({\bf r_4},t^{\prime}) \rangle_c
=\langle \Psi({\bf r_1},t)\Psi^*({\bf r_4},t^{\prime})\rangle
\langle \Psi({\bf r_3},t^{\prime})\Psi^*({\bf r_2},t) \rangle, 
\label{Wick}
\end{equation}
retaining only the connected piece.  
 The Gaussian term is represented diagrammatically in Figure 
\ref{Gaussian}.
 An arrow corresponds to a Green's function, and a circle 
corresponds to the noise average.  
 The combination arrow-circle-arrow constitutes a correlation 
function. 
 Hence the conductivity at Gaussian order requires two correlation 
functions. 
 Not drawn but of crucial importance are the momentum operators 
acting at sites $({\bf r},t)$ and $({\bf r^{\prime}},t^{\prime})$ 
that make it a current-current correlator instead of a 
density-density correlator.  

Let us focus our attention for now on the uniform dc conductivities 
$\sigma_{xx}({\bf 0})$ and $\sigma_{zz}({\bf 0})$. 
 Calculated in the LLL approximation and to Gaussian order they are 
\begin{equation}
\sigma_{xx}^{(G)}({\bf 0}) = {e^{*2} ~\langle |\Psi_0|^2 \rangle 
\over m_{ab}  } \left({1 \over \Gamma \hbar \omega_0 } \right)
\label{uniform-xx-2}
\end{equation}
and
\begin{equation}
\sigma_{zz}^{(G)}({\bf 0}) =  {e^{*2} ~\langle |\Psi_0|^2 \rangle 
\over m_c} \left({1 \over 8 \Gamma \alpha_H  } \right)
\label{uniform-zz-2}
\end{equation}  
 We see here the characteristic time of the $ab$-plane 
conductivity is $\tau_{ab}=1/\Gamma \hbar \omega_0$ while 
that corresponding to the $c$-axis conductivity is $\tau_c=1/8 
\Gamma \alpha_H$. 

Consider now how the Kubo formula reproduced the flux-flow result 
at Gaussian order. 
 First of all the momentum operators intrinsic to $\sigma_{xx}
({\bf 0})$ act like creation or annihilation operators raising or 
lowering the Landau level with the result that one of the $\Psi$'s 
at $({\bf r},t)$ is in a higher level; the same thing happening at 
$({\bf r^{\prime}},t^{\prime})$ as well.   
 Consequently, in the LLL limit, we have one $n=0$ correlator and 
one $n=1$ correlator.  
 With this in mind, we now look at the time integral in the Kubo 
formula, which becomes 
\begin{equation}
\int d(t-t^{\prime}) ~{\rm exp} \left\{ -\Gamma \left[E_0(k_{z1}) 
+ E_1(k_{z2})\right]|t-t^{\prime}| \right\}, 
\end{equation} 
where $k_{z1}$ and $k_{z2}$ are the $z$ components of momentum 
running through the $n=0$ and $n=1$ channels, respectively.
 (Actually $k_{z1}=k_{z2}$ by momentum conservation.) 
 Because $E_0(k_{z1})$ is much smaller than $E_1(k_{z2})$ we can 
drop $E_0(k_{z1})$ from the integrand above. 
 This last step is equivalent to using a static (equilibrium) 
$n=0$ correlator and hence produces a relation between the 
dynamic conductivity and the static density.
  In fact, the integral yields simply a factor $1/\Gamma \hbar 
\omega_0$ --- the characteristic time in the flux-flow formula. 
 The flux-flow formula obtained in the direct approach above 
suggests that the Gaussian density $\langle |\Psi_0|^2 \rangle$ 
should be replaced by the fully renormalized density $\langle |\tilde 
\Psi_0 |^2 \rangle$ but that the characteristic time $\tau_{ab}=
1/\Gamma \hbar \omega_0$ should remain as is. 
 The way to produce this result within the Kubo formalism is to 
renormalize the $n=0$ correlator while leaving the $n=1$ correlator 
essentially untouched.
 If this is true then diagrams that dress up or otherwise disrupt the 
bare $n=1$ correlator should have no overall effect. 

\subsection{The Hartree-Fock Approximation}

Our first attack on going beyond Gaussian order, {\it i.e.} 
incorporating the nonlinear terms, will be the Hartree-Fock 
approximation.  
 At Gaussian order the superconducting fluctuations are 
\begin{equation}
\langle |\Psi_0|^2 \rangle = {k_BT \over 4 \pi ~ \alpha_H ~ 
\xi_c \ell^2} ,
\label{density}
\end{equation} 
which would seem to imply that the conductivity diverges at 
mean-field $H_{c2}$ ($\alpha_H=0$). 
 This seeming divergence is eliminated when one adopts the 
Hartree or Hartree-Fock approximation. \cite{Ullah,Ikeda2}
 
\begin{figure}
\centerline{
\epsfxsize=7cm \leavevmode \epsfbox{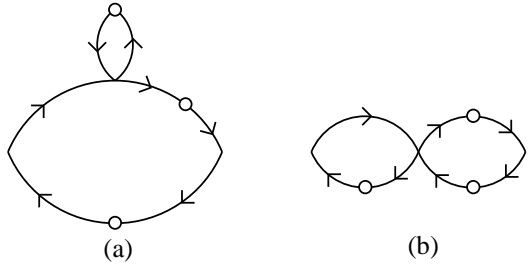}}
\caption{ Diagrams with one quartic vertex. Diagram 
\ref{order-b}(a) is included in the Hartree-Fock resummation. 
Diagram \ref{order-b}(b) has zero contribution at ${\bf k=0}$ 
and is ``down" in the calculation of $\sigma_{xx}({\bf k 
\neq 0})$.}  
\label{order-b}
\end{figure}

From the perturbation theory for $\Psi$ (\ref{iterate-b}) we 
can obtain the series for $\langle \Psi({\bf r_1},t)
\Psi^*({\bf r_2},t)\Psi({\bf r_3},t^{\prime})\Psi^*(
{\bf r_4},t^{\prime}) \rangle$ and in turn the series for the 
conductivity. 
 The new diagrammatic feature is a ``quartic" vertex at which 
four lines meet with two arrows pointing in and two pointing out. 
 Two diagrams occurring at the first order in $\beta$ are shown 
in Figure \ref{order-b}.  
 Because of the structure of the conductivity, especially the 
momentum operators acting at $({\bf r},t)$ and $({\bf r^{\prime}}, 
t^{\prime})$, the diagram shown in Fig. \ref{order-b}(b) has zero 
contribution at ${\bf k=0}$ for both $\sigma_{xx}$ and 
$\sigma_{zz}$.   

\begin{figure}
\centerline{
\epsfxsize=7cm \leavevmode \epsfbox{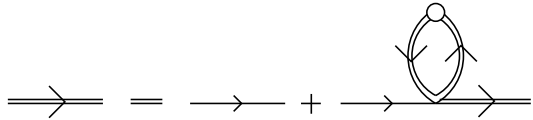}}
\caption{ Diagrammatic representation of the Hartree-Fock 
approximation. The double lines represent renormalized 
Green's functions; the single lines represent bare Green's 
functions.} 
\label{HF-fig}
\end{figure}

 Diagrams like that in Fig. \ref{order-b}(a) can be resummed 
by replacing $\alpha_H$ with $\tilde \alpha$, where $\tilde 
\alpha$ is defined self-consistently as $\tilde \alpha = 
\alpha_H +2 \beta \langle |\tilde \Psi_0|^2 \rangle$, where 
what we mean here by $\langle |\tilde \Psi_0|^2 \rangle$ is 
$\langle | \Psi_0|^2 \rangle$ (eq. (\ref{density}) with 
$\alpha_H \rightarrow \tilde \alpha$; all of which leads to
\begin{equation}
\tilde \alpha = \alpha_H + {\beta ~k_B T  \over 
2 \pi ~\ell^2  \xi_c  ~\tilde \alpha}.  
\label{HF}
\end{equation}
 Replacing $\alpha_H$ with $\tilde \alpha$ constitutes a 
resummation, the Dyson equation for which is represented 
diagrammatically in Fig. \ref{HF-fig}. 
 It corresponds to a renormalization of the Green's function. 

In analogy with many-body physics, this replacement is called 
the Hartree-Fock (HF) approximation as it takes into account 
both the direct and exchange terms. 
 In addition to eliminating a large number of graphs from the 
perturbation theory; the HF resummation replaces $\alpha_H$ 
which can be zero or negative with $\tilde \alpha$ which cannot 
--- curing the divergence problem mentioned above. 
 The removal of the divergence is connected to the fact that 
a large magnetic field effectively reduces the dimension of the 
problem by two. \cite{Ruggeri}
 Moreover, the HF approximation goes a long way toward achieving 
agreement with measured conductivities. \cite{screen} 
 This success suggests that the same philosophy ($\alpha_H 
\rightarrow \tilde \alpha$) should be adopted when considering 
where the LLL approximation is valid. 
 In the HF approximation $\alpha_H \rightarrow \tilde \alpha$ 
while $\hbar \omega_0$ remains the same. 
 So whereas it originally appeared that the LLL approximation 
was valid near the mean-field $H_{c2}(T)$ line where $\alpha_H$ 
is small, it would now appear to be valid where $\tilde \alpha$ 
is small. 
 But as already noted $\tilde \alpha$ does not change sign, it 
grows small only as $\alpha_H$ grows large and negative, {\it i.e.} 
below the mean-field $H_{c2}(T)$ line. 
 Ikeda \cite{Ikeda1} has investigated renormalization effects 
beyond HF, finding that they are merely refinements to the HF 
considerations, suggesting that the LLL approximation has a 
substantial region of validity within the vortex-liquid regime.  

If we replace $\alpha_H$ with $\tilde \alpha$ in the Gaussian 
conductivities and look in the limit of $\alpha_H \ll 0$ we find 
that  
\begin{equation}
\lim_{\alpha_H \ll 0}\ \sigma_{xx}({\bf 0}) = {e^* |\alpha_H| 
\over 2 \beta \hbar B \Gamma} .
\end{equation}
Note that $\sigma_{xx}({\bf 0})$ at this level of approximation 
already shows many of the features seen in the experiments in the 
vortex-liquid regime: the {\it resistivity} is linear in $B$ with 
a zero intercept and extrapolates to the normal resistivity at the 
mean-field $H_{c2}(T)$ line, as can be seen for instance in the 
Fendrich {\it et al.} data prior to irradiation.  \cite{Fendrich}

\subsection{Beyond the Hartree-Fock Approximation}

While the Hartree or HF approximation goes along way toward 
describing certain properties such as the specific heat, it is 
totally inadequate for examining other features, such as the 
extent of crystalline ordering within the $ab$ plane. 
 A much more sophisticated approach, such as the Parquet 
resummation \cite{Yeo}, is needed for that. 
 We do not provide such a scheme here, instead we examine a 
few diagrams beyond the HF approximation. 
 We will find a cancellation among most of these diagrams at 
${\bf k=0}$ suggesting that the Hartree approximation is 
already quite good for the uniform conductivity. 
 The absence of this cancellation at ${\bf k \neq 0}$, on 
the other hand, suggests the need to go beyond Hartree theory 
when considering the nonlocal effects.  

As we are now considering a perturbation theory about the HF 
approximation, the series is no longer a power series in $\beta$. 
 Rather it is a power series in $x$ where 
\begin{equation}
x = {\beta k_BT  \over 
16 \pi \ell^2 \xi_c {\tilde \alpha}^{2}}, 
\end{equation}
the dimensionless parameter introduced by Ruggeri and 
Thouless. \cite{Ruggeri}
 The self-consistent equation above, eq. (\ref{HF}), is now rather 
compactly written as 
\begin{equation}
\alpha_H = \tilde \alpha (1-8x).
\end{equation}

\begin{figure}
\centerline{
\epsfxsize=7cm \leavevmode \epsfbox{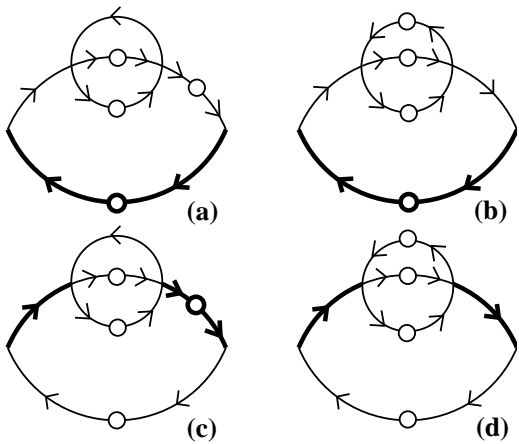}}
\caption{ Diagrams contributing to $\sigma_{xx}$ at order 
$x^2$. The bold lines indicate $n\geq 1$ Green's functions; 
the thin lines $n=0$ Green's functions.  Diagrams (a) and 
(c) are Green's function renormalizing; while (b) and (d) are 
noise-renormalizing --- the difference being in the placement 
of the circles (noise averages). Diagrams (a) and (b) 
contribute to the flux-flow formula, (c) is ``down" and does 
not contribute to the LLL approximation, and (d) is canceled 
by another diagram at ${\bf k=0}$. }  
\label{x2-fig-1}
\end{figure}

\begin{figure}
\centerline{
\epsfxsize=7cm \leavevmode \epsfbox{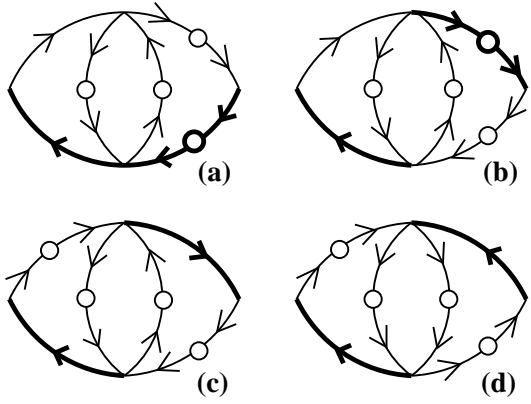}}
\caption{ Vertex renormalizing diagrams contributing to 
$\sigma_{xx}$ at order $x^2$. 
Diagrams (a) and (b) are ``down;" diagram (c) is zero at 
${\bf k=0}$; and (d) is canceled by the diagram in Fig. 
\ref{x2-fig-1} (d). }  
\label{x2-fig-2}
\end{figure}

\begin{figure}
\centerline{
\epsfxsize=7cm \leavevmode \epsfbox{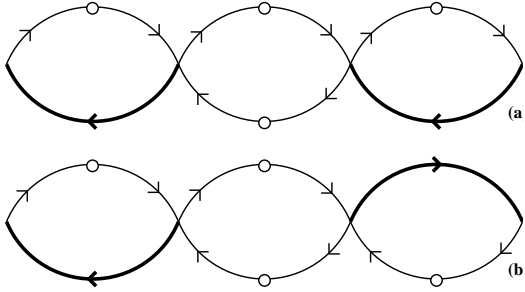}}
\caption{ More vertex renormalizing diagrams contributing to 
$\sigma_{xx}$ at order $x^2$. 
These diagrams do not contribute at ${\bf k=0}$. }  
\label{x2-fig-3}
\end{figure}

In the expansion of the uniform conductivity $\sigma_{xx}
({\bf 0})$ there are no contributions of $O(x)$. 
 Recall that of the diagrams at order $\beta$ (Fig. 
\ref{order-b}) there are those of the HF type which have been 
absorbed into $\tilde \alpha$, and the others make no 
contribution at ${\bf k=0}$. 
 Some diagrams contributing to the conductivity at $O(x^2)$ 
are shown in Figures \ref{x2-fig-1}, \ref{x2-fig-2} and 
\ref{x2-fig-3}. 

There are a number of considerations in addition to the usual 
ones of the topology and degeneracy to keep in mind when 
enumerating and evaluating the diagrams that occur in the 
perturbation series for $\sigma_{xx}$.  
 Because the conductivity is a dynamic quantity, there is the 
distinction between Green's functions and correlation 
functions to bear in mind.  
 We make this differentiation by including a small circle 
representing the noise average in the middle of a correlation 
function.  
 Diagrams may differ in the placement of circles; for example, 
compare Figs. \ref{x2-fig-1}(a) and (b) or Figs 
\ref{x2-fig-2}(b) and (c).
 In addition, because the order parameter is complex, the 
Green's function carries a direction, indicated by the arrow. 
 Hence diagrams may be distinguished by the placement of arrows 
around the diagram; for instance, compare Figs. 
\ref{x2-fig-2}(c) and (d). 
 Also because the order parameter is expanded in Landau levels, 
the Landau level of each line is another consideration.  
 We make this differentiation by putting the higher Landau-level 
Green's functions in bold. 
 So diagrams may differ in Landau level structure; for example, 
compare Figs. \ref{x2-fig-1}(a) and (c) or Figs. 
\ref{x2-fig-2}(a) and (b).  On the other hand, in the evaluation 
of $\sigma_{zz}({\bf 0})$ all of the Green's functions are in 
the LLL. 

 Recall that at Gaussian order $\sigma_{xx}({\bf 0})$ has one 
correlation function and thus two Green's functions in the 
$n=1$ state. 
 It turns out that at all orders in perturbation theory the 
diagrams contributing to the LLL approximation have a maximum 
of two higher-Landau-level Green's functions.
 So diagrams like that in Fig. \ref{x2-fig-1}(c) and those in 
Figs. \ref{x2-fig-2} (a) and (b) are ``down" by a factor of 
$\tilde \alpha / \hbar \omega_0$ and are not included to this 
order of approximation. 
 Of the remaining diagrams it has been argued that those which 
have a simple $n=1$ correlator ({\it e.g.} Figs. \ref{x2-fig-1}(a) 
and (b)) contribute to the flux-flow result and that the others 
({\it e.g.} Figs. \ref{x2-fig-1}(d), \ref{x2-fig-2}(c) and (d) 
and \ref{x2-fig-3}(a) and (b)) must have a combined contribution 
of zero at ${\bf k=0}$.  
 When we evaluated these diagrams, we indeed found that the 
diagram in Figs. \ref{x2-fig-2}(c), \ref{x2-fig-3}(a) and 
\ref{x2-fig-3}(b) gave no contribution at ${\bf k=0}$ and that 
the contributions due to the diagrams in Figs. \ref{x2-fig-1}(d) 
and Figs. \ref{x2-fig-2}(d) were equal and opposite.   
 We will demonstrate this explicitly in the $2D$ case. 

 The diagrams in Fig. \ref{x2-fig-1}(a) and (c) renormalize 
the Green's function, and those in Fig. \ref{x2-fig-1}(b) and 
(d) renormalize the noise. 
 The diagrams in Figs. \ref{x2-fig-2} and \ref{x2-fig-3} 
renormalize the quartic vertex.   
 Note that on the $n=1$ side, the Green's function renormalizing 
diagram was down whereas the noise renormalizing diagram was not. 
 A similar phenomenon occurs when we consider the effect of 
disorder.    

\section{Wave-vector dependence}

Now let us turn to the nonlocal, ${\bf k}$-dependent, 
conductivity.  
 In some instances we calculate the full ${\bf k}$-dependence 
\cite{BM}, while in others we restrict our attention to 
the coefficients in a small-${\bf k}$ expansion
\begin{equation}
\sigma_{\mu \nu}({\bf k}) \ = \ \sigma_{\mu \nu}({\bf 0})
\ + \ S_{\mu \alpha \beta \nu}~k_{\alpha}k_{\beta} + O(k^4). 
\label{hydro-notation}
\end{equation}
 The coefficients $S_{\mu \alpha \beta \nu}$ and the 
associated length scales can be related to  the vortex 
picture.  
  Since the vortices move in the $y$ direction when the 
electric field and current are in the $x$ direction, 
$S_{xxxx}$ is associated with shearing \cite{Mouetal} and 
$S_{xyyx}$ with compression. 
 $S_{xzzx}$ is related to the ``integrity" of vortices 
(Huse and Majumdar\cite{HusMaj} call it the ``tilt 
viscosity"). 

\subsection{The Gaussian results}
  
 The small wave-vector expansion of the conductivity at 
Gaussian order (within the combined HF and LLL 
approximations) is 
\begin{equation}
\sigma_{xx}^{(G)}({\bf k}) =  \sigma_{xx}^{(G)}({\bf 0})
\left[ 1 - {\ell^2 k_x^2 \over 4} + 
{3 \hbar^2 \omega_0^2 \ell^2 k_y^2 \over 64 {\tilde 
\alpha}^2} - \ell_c^2 k_z^2 + O(k^4) \right], 
\label{gauss-k}
\end{equation}
where we have factored out $\sigma_{xx}^{(G)}({\bf 0})$. 
 First note that $S_{xxxx}$ and $S_{xzzx}$ are negative, 
{\it i.e.} they have the sign that cannot be handled using 
the hydrodynamic approach. 
 Furthermore, the length scales multiplying $k_y^2$ and 
$k_z^2$ are magnetic lengths, {\it i.e.} they have no 
temperature dependence. 

It is easy to see why (calculationally) the $c$-axis length 
scale in the Gaussian calculation of $\sigma_{xx}^{(G)}$ is 
$\ell_c$.  
 The $z$-dependence has no effect on the Landau level structure; 
therefore, it remains true that $\sigma_{xx}(k_z)$ is comprised 
of one $n=0$ and one $n=1$ correlator.  
  We can take advantage of the disparity in ``masses" ($\tilde 
\alpha \ll \hbar \omega_0$) by insisting that the external 
momentum $k_z$ be sent through the $n=1$ channel.  
 (This choice does not affect the outcome, it just makes it 
more apparent.)  
 The internal momentum integral is dominated by the smallest 
poles, which are those associated with $\tilde \alpha$, and 
the terms involving $k_z$  and $\hbar \omega_0$ are left 
essentially intact.  
The full $k_z$ dependence at Gaussian order is thus 
\begin{equation}
\sigma_{xx}^{(G)}(k_z) = \sigma_{xx}^{(G)}({\bf 0}) 
\left[ 1+k_z^2 \ell_c^2/2 \right]^{-2},
\label{gauss-kz} 
\end{equation} 
where the structure is simply that of $[E_1(k_z)]^{-2}$ 
with $\tilde \alpha \rightarrow 0$.  

The Landau level structure of $S_{xxxx}$ is slightly different.  
 This time one correlator is in the $n=0$ level while the 
other is in either the $n=1$ or the $n=2$ level.  
 For an expansion in $k_x^2$ each additional power of $k_x^2$ 
requires one higher Landau level.   
 But the important point is that there is no contribution 
with two $n=0$ correlators. 
  The $k_x$ dependence at Gaussian order is 
\begin{equation}
\sigma_{xx}^{(G)}(k_x) = \sigma_{xx}^{(G)}({\bf 0}) \left[ 
{ 1-e^{-k_x^2 \ell^2/2} \over k_x^2 \ell^2/2}\right].
\label{gauss-kx} 
\end{equation} 
The combined $k_x$ and $k_z$ dependences are given by
\begin{equation}
\sigma_{xx}^{(G)}(k_x,k_z) = \sigma_{xx}^{(G)}({\bf 0}) 
\left[
{\rm e}^{- k_x^2 \ell^2 / 2 }
 \sum_{n=0}^{\infty} {(n+1)\left(k_x^2 \ell^2 / 2\right)^n 
\over n! \left[ (n+1) + k_z^2 \ell_c^2/ 2 \right]^2} 
\right].
\label{k-depend}
\end{equation}

 As opposed to $S_{xxxx}$ and $S_{xzzx}$, $S_{xyyx}$ is 
positive at Gaussian order. 
 In fact if we had included the next term in the $\tilde 
\alpha / \hbar \omega_0$ expansion we would see that 
$S_{xyyx}$ changes sign as $T$ is lowered. \cite{Mouetal}  
 As $T$ decreases the associated length scale associated 
with $S_{xyyx}$, $\xi_{\perp}\propto \hbar \omega_0 
\ell/\tilde \alpha$, increases. 
 A $T$-dependent length scale here is somewhat surprising. 
 In the vortex picture $S_{xyyx}$ appears to be related 
to compression, but one might expect that the compressibility 
to be pretty much the same for the liquid and the solid, and 
so relatively $T$-independent. 
 The Gaussian calculation is in conflict with this expectation. 
 If this length scale found in the transverse conductivity 
does indeed increase as $T$ decreases, it will prove 
interesting to compare it to similar growing $ab$-plane length 
scales, for example, the phase coherence length and the length 
over which density-density fluctuations decay \cite{Mike}, which 
recent Monte Carlo simulations suggest grow in the same 
way\cite{Matthew}.

In the scenario in which we apply current and wish to extract 
the voltage distribution, it is the characteristic lengths of 
the resistivity which are important. 
 This is why the signs of the $S$'s are so crucial. 
 They determine the pole structure of the nonlocal resistivity 
$\rho({\bf k})$ and consequently the length scales of 
$\rho({\bf r})$. 
 Consider, as an example, a $c$-axis conductivity of the form 
\begin{equation}
\sigma_{zz}(k_z) = \sigma_{zz}^{(n)}+ \sigma_{zz}^{(s)}(0) 
+ S_{zzzz}k_z^2 + O(k^4), 
\end{equation}
where a local normal conductivity has been included. 
 If $S$ is positive, the hydrodynamic approach \cite{HusMaj} 
yields a length scale $[S/(\sigma^{(s)}+\sigma^{(n)})]^{1/2}$. 
 If $S$ is negative, a Pad\'e approximant approach \cite{BM} 
produces a length scale $[\sigma^{(n)} |S|/\sigma^{(s)}
(\sigma^{(s)}+\sigma^{(n)})]^{1/2}$ which is much smaller than 
the previous one because of the factor $(\sigma^{(n)}/
\sigma^{(s)})^{1/2}$. 
 Returning to the case of $S_{xyyx}$, if it were to change 
sign from positive to negative as more diagrams are included, 
the associated length scale would then be short which may be 
consistent with the expectations of the vortex picture.   
 With such dramatic consequences regarding the nonlocal 
behavior of the resistivity, one might expect that changes in 
the signs of the $S$'s would have experimental signatures.  

 The difference between $S_{xyyx}$ and the others is that all 
four $\Psi$'s in the calculation of $S_{xyyx}$ can be in the 
LLL.   
 Applying the definition of $S_{xyyx}$, namely 
\begin{equation}
S_{xyyx} = {1 \over 2}{\partial^2 \sigma_{xx}({\bf k}) 
\over \partial k_y^2 } \Biggl|_{{\bf k=0}},
\end{equation}
to the Kubo formula (eq. (\ref{kubo2})) and restricting to 
LLL states yields
\begin{eqnarray}
S_{xyyx} &\ = \ &-{e^{*2} \over 16 k_B T m_{ab}^2 }
\int d({\bf r}-{\bf r^{\prime}}) \int d(t-t^{\prime}) 
~(y-y^{\prime})^2 \Bigl(P_{1x}+P_{2x}^*\Bigr)
\Bigl(P_{3x}+P_{4x}^*\Bigr) 
\nonumber \\ 
&&\ \ \  
\langle  \Psi_0({\bf r_1},t) \Psi_0^*({\bf r_2},t) 
\Psi_0({\bf r_3},t^{\prime})\Psi_0^*({\bf r_4},t^{\prime})
\rangle_c \Biggl|_{{\bf r_1=r_2=r} \atop 
{\bf r_3=r_4=r^{\prime}}} .
\label{Monte-1}
\end{eqnarray}
 This expression involves only LLL states is hence accessible 
to simulations and other methods which use only LLL states.  
\cite{Anne} 

Another quantity that involves only LLL states is the $c$-axis 
conductivity, $\sigma_{zz}({\bf k})$, at Gaussian order it is 
\begin{equation}
\sigma^{(G)}_{zz}({\bf k}) = \sigma_{zz}^{(G)}({\bf 0}) 
\left[ {\rm e}^{k_{\perp}^2\ell^2/2} \left({1-(1+k_z^2 
\xi_c^2/4)^{-1/2} \over k_z^2\xi_c^2/8} \right) \right],
\end{equation}
where $k_{\perp}^2=k_x^2+k_y^2$. 
 Since the calculation involves only $n=0$ states, the 
$c$-axis length scale is necessarily $\xi_c$. 
 If this result is expanded one finds that none of $S$'s 
has the sign required by the hydrodynamic approach. 

\subsection{Beyond Gaussian}

  As already mentioned the HF approximation is inadequate 
for determining the extent of crystal ordering within the 
$ab$ plane and as such is also inadequate for examining 
shear effects.  
 To see if $S_{xxxx}$ has a long length scale or changes 
sign at low $T$, we must look beyond the HF approximation.  
 The same is true of $S_{xzzx}$. 
 
We argued that the uniform conductivity, $\sigma_{xx}
({\bf k=0})$, could be obtained from the subset of diagrams 
that renormalize the $n=0$ correlator and leave the $n=1$ 
untouched beyond the HF resummation.  
 In the LLL limit the fully renormalized $n=0$ correlator 
$\tilde C_{n=0}({\bf r}, t;{\bf r^{\prime}},t^{\prime})$ has 
{\it exactly} the same dependence on $x$, $x^{\prime}$, $y$ 
and $y^{\prime}$ as its ``bare" version; thus, the $k_x$-dependence 
of this particular resummation is exactly the same as that 
in the Gaussian calculation (eq. (\ref{gauss-kx})).  
 Furthermore, the external momentum $k_z$ can still be sent 
through the $n=1$ channel, yielding again the dependence seen 
in the Gaussian calculation (eq. (\ref{gauss-kz})). 
 Resumming this subset of diagrams only alters the magnitude 
$\sigma_{xx}({\bf 0})$; the $k_x$ and $k_z$ dependence remain 
the same as they were in the Gaussian calculation, {\it i.e.} 
\begin{equation}
\sigma_{xx}^{(FF)}(k_x,k_z) = \sigma_{xx}({\bf 0}) 
\left[
{\rm e}^{- k_x^2 \ell^2 / 2 }
 \sum_{n=0}^{\infty} {(n+1)\left(k_x^2 \ell^2 / 2\right)^n 
\over n! \left[ (n+1) + k_z^2 \ell_c^2/ 2 \right]^2} \right].
\end{equation} 
where 
\begin{equation}
\sigma_{xx}({\bf 0})= {e^{*2} \langle |\tilde \Psi_0 |^2 
\rangle \over m_{ab} \Gamma \hbar \omega_0}
\end{equation}
and where 
\begin{equation}
\langle |\tilde \Psi_0 |^2 \rangle = {k_BT \over 4 \pi ~ 
\tilde \alpha ~ \xi_c \ell^2} \left[ 1 + O(x^2) \right]. 
\end{equation}
So for this particular subset of diagrams, we only have to 
do the perturbative expansion for the static quantity 
$\langle |\tilde \Psi_0 |^2 \rangle$. 
 Consequently, if $S_{xxxx}$ and $S_{xzzx}$ are going to change 
sign at low $T$ and give long length scales, then other diagrams, 
such as the vertex-renormalizing diagrams (see Fig. \ref{x2-fig-2}), 
must be important. 

 Recall that in the $\tilde \alpha \ll \hbar \omega_o$ limit 
we only want diagrams with at most two higher Landau level 
($n\geq1$) Green's functions. 
 Green's functions meeting at a noise average must be in the 
same Landau level; while those meeting at a vertex can be in 
different Landau levels. 
 So either the two $n \geq1$ Green's functions meet each other 
at a noise average, or each meets three $n=0$ Green's functions 
at a vertex.  
 Otherwise, there will be at least three higher Landau level 
Green's functions; see for example, Fig. \ref{x2-fig-1}(c). 
 We have already considered those diagrams in which the two 
$n\geq1$ Green's functions meet each other at a noise average 
--- they are the ones that yield the flux-flow result 
$\sigma^{(FF)}_{xx}({\bf k})$ --- they are also essentially 
two-point quantities and thus do not probe viscous effects such 
as shearing.  
 
Let us find an expression for those terms in which each 
$n\geq 1$ Green's functions meets a vertex. 
 Recall the TDGL equation in symbolic form $\Psi_1 = G_{1,2} 
\eta_2 -\beta G_{1,2}\Psi^*_2 \Psi_2 \Psi_2$ (eq. (\ref{symbol})). 
 Of the four $\Psi$'s in the Kubo formula, let us replace the 
two that are in higher Landau levels with the second term on 
the right hand side of eq. (\ref{symbol}).  
 This substitution yields    
\begin{eqnarray}
\Sigma_{xx}(k_x,k_z) &=& {e^{*2} \beta^2\over 8 k_B T m_{ab}^2}
\int d({\bf r}-{\bf r^{\prime}}) \int d(t-t^{\prime}) 
\int_5 \int_6~{\rm e}^{i k_x (x- x^{\prime})+ 
i k_z (z- z^{\prime}) } \nonumber \\ 
&&\Bigl(P_{1x}+P_{2x}^*\Bigr)\Bigl(P_{3x}+P_{4x}^*\Bigr)  
\Biggl\{ G_{2,5}^*G_{3,6} \langle \Psi_1 \Psi_5^* 
\Psi_5^* \Psi_5 \Psi_6 \Psi_6 \Psi_6^* \Psi_4^* 
\rangle_c \nonumber \\
&&\ \ \ \ \ \ \ \  + G_{2,5}^*G_{4,6}^*  \langle \Psi_1 
\Psi_5^* \Psi_5^* \Psi_5 
\Psi_6^* \Psi_6^* \Psi_6 \Psi_3 \rangle_c  + c.c. \Biggr\} 
 \Biggl|_{1=2=({\bf r},t) \atop 
3=4=({\bf r^{\prime}},t^{\prime})} .
\label{not-flux-flow}
\end{eqnarray}      
In this symbolic notation, the numbers denote points in 
space-time, {\it e.g.} $5 \rightarrow ({\bf r_5},t_5)$. 
 This expression factors out the ``bare" $n \geq 1$ Green's 
functions explicitly, leaving averages which involve LLL 
states only. 
 The LLL average is a rather complicated eight-point object, 
but in principle, coupled with eqs. (\ref{k-depend}) and 
(\ref{Monte-1}), it allows one to find the conductivity 
from simulations that use only LLL states.  
 Because the Green's functions that have been factored out 
involve higher Landau level states, they are rather short-ranged 
in both time and space. 
 When investigating a regime in which one expects the LLL 
object to become long-ranged, it is tempting to approximate 
the Green's function by some appropriately scaled delta 
function $\delta ({\bf r}- {\bf r^{\prime}}) 
\delta(t-t^{\prime})$ after the derivatives have acted.    

 In the next section we will provide some explicit 
expressions for the diagrams in Figs. \ref{x2-fig-1}, 
\ref{x2-fig-2} and \ref{x2-fig-3} for the two-dimensional 
case. 
 But before leaving this section, let us make one point 
about the $k_z$ behavior.  
 As previously mentioned the diagrams \ref{x2-fig-1}(d) 
and \ref{x2-fig-2}(d) cancel at $k_z=0$, but note that in 
these diagrams there is no longer an $n=1$ channel through 
which we can pass the external momentum $k_z$. 
 Therefore, one might expect that one picks up the $\xi_c$ 
length scale instead of $\ell_c$, however, this is not the 
case. 
 Just as these diagrams cancel at ${\bf k=0}$ so do the 
leading terms ({\it i.e.} the $k_z^2 \xi_c^2$ terms) in 
the $k_z$ expansion. 
  
\section{Two-dimensional results}
 
 Many of the results above hold for films and even simplify 
in this case.  
 The uniform conductivity is still given by the flux-flow 
formula in the LLL approximation. 
 The diagrammatic structure of the perturbation series is 
exactly the same; and which diagrams are down and so forth 
also carries through; only the evaluation of the diagrams 
changes.    

To do these calculations we will of course need the $2D$ 
Green's function, which in the absence of an electric field is
\begin{equation}
G_{2D}({\bf r},t;{\bf r^{\prime}}, t^{\prime}) = 
{\Gamma \over L_z}\int{d \omega \over 2 \pi} \int 
{d k_y \over 2 \pi}   {\rm e}^{-i\omega(t-t^{\prime}) 
+ik_y(y-y^{\prime}) }  \sum_{n=0}^{\infty}{ u_n\left(
{x \over \ell} -k_y \ell \right)
u_n\left({x^{\prime} \over \ell} -k_y \ell \right) \over 
\Gamma E_n -i\omega }, 
\label{green-2d-1}
\end{equation}
where $E_n= \alpha + \hbar \omega_0 (n + 1/ 2)$ and $L_z$ 
is the film thickness. 
 The Gaussian result for the conductivity is 
\begin{equation}
\sigma_{xx} ({\bf k})= {e^{*2} \langle |\tilde \Psi_0|^2 
\rangle_{2D} \over m_{ab} \Gamma \hbar \omega_0}  
\left[ 1- {\ell^2 k_x^2 \over 4}  + { \hbar^2 \omega_0^2 
\ell^2 k_y^2  \over 8 {\tilde \alpha_{2D}}^2} +O(k^4)
 \right]
\label{2D-gauss}
\end{equation}
in the LLL limit.  
 Other than a different numerical coefficient in front of 
the $k_y^2$ ($S_{xyyx}$), the result is similar to the $3D$ 
result (eq. (\ref{gauss-k})). 
 Again we are applying the HF approximation so that $\tilde 
\alpha = \alpha_H + 2 \beta \langle |\tilde \Psi_0|^2 \rangle$ 
where 
\begin{equation}
\langle |\tilde \Psi_0|^2 \rangle_{2D} = {k_BT \over 2 \pi ~ 
\tilde \alpha_{2D} ~ L_z \ell^2},
\label{density-2D}
\end{equation}
which leads to the self-consistent equation 
\begin{equation}
\tilde \alpha_{2D} = \alpha_H - 
{\beta ~k_BT \over \pi ~\ell^2 L_Z ~\tilde \alpha_{2D}}.
\label{HF2D}
\end{equation}
The expansion parameter in the two-dimensional perturbation 
series is 
\begin{equation}
x_{2D} = {\beta ~k_BT \over 4 \pi ~\ell^2 L_z ~{\tilde 
\alpha_{2D}}^2}, 
\end{equation}
allowing one to re-express eq. (\ref{HF2D}) as $\alpha_H= 
\tilde \alpha_{2D} (1-4 x_{2D})$. \cite{Ruggeri}

Evaluating some of the diagrams in Figs. \ref{x2-fig-1} 
and \ref{x2-fig-2} gives
\begin{eqnarray}
\sigma_{xx}^{(FF)}&=& \sigma_{xx}^{(G)}({\bf 0}) 
~{7x_{2D}^2 \over 2} ~\left[{\rm e}^{-k_{\perp}^2\ell^2/2}
\left( 1- k_y^2\ell^2+{1 \over 4}k_y^2k_{\perp}^2\ell^4 
\right) \right]
\\
\sigma_{xx}^{4(d)}&=& \sigma_{xx}^{(G)}({\bf 0}) 
~{x_{2D}^2 \over 4} ~\left[{\rm e}^{-k_{\perp}^2\ell^2/2}
\left( 1- k_y^2\ell^2+{1 \over 4}k_y^4\ell^4 
+k_x^2k_y^2\ell^4 \right) \right]
\\
\sigma_{xx}^{5(c)}&=& \sigma_{xx}^{(G)}({\bf 0}) 
~{x_{2D}^2 \over 8} ~\left[{\rm e}^{-3k_{\perp}^2\ell^2/4}
\left( -k_x^2\ell^2+k_y^2\ell^2 -k_y^2k_{\perp}^2\ell^4 
+{1 \over 4} k_y^2k_{\perp}^4 \ell^6\right) \right]
\\
\sigma_{xx}^{5(d)} &=& \sigma_{xx}^{(G)}({\bf 0}) 
~{x_{2D}^2 \over 4} ~\left[ {\rm e}^{-k_{\perp}^2\ell^2}
\left( -1+k_x^2\ell^2+2k_y^2\ell^2-{5 \over 4}
 k_y^2k_{\perp}^2 \ell^4 
+{1 \over 4}k_y^2k_{\perp}^4 \ell^6 \right) \right]
\\
\sigma_{xx}^{6(a)} &=& \sigma_{xx}^{(G)}({\bf 0}) 
~{2x_{2D}^2 } ~\left[ {\rm e}^{-3k_{\perp}^2\ell^2/2}
\left( k_{\perp}^2\ell^2-k_y^2k_{\perp}^2\ell^4+
{1 \over 4} k_y^2k_{\perp}^4 \ell^6  
\right) \right]
\\
\sigma_{xx}^{6(b)} &=& \sigma_{xx}^{(G)}({\bf 0}) 
~{2x_{2D}^2 } ~\left[ {\rm e}^{-3k_{\perp}^2\ell^2/2}
\left( -k_x^2\ell^2+k_y^2\ell^2-k_y^2k_{\perp}^2\ell^4
+{1 \over 4} k_y^2 k_{\perp}^4\ell^6  \right) \right].
\end{eqnarray}
These calculations were done with the Green's function 
shown in bold in the figures in the $n=1$ state.  
 Recall that if we are interested in $S_{xyyx}$ we must 
redo the calculation with all $n=0$ states, and if we are 
interested in $S_{xxxx}$ we must also add to the expressions 
above those with $n=2$ Green's functions. 
 Here our interest is in showing explicitly the overall 
null contribution of these diagrams at ${\bf k=0}$ --- 
whether it be that the individual diagrams have zero 
contribution ({\it e.g.} diagrams 5(c), 6(a) and 6(b)) or 
that diagrams cancel ({\it e.g.} diagrams 4(d) and 5(d)) 
--- and to suggest that the same does not apply at 
${\bf k \neq 0}$.  

\section{ Disorder}

Disorder introduces competing effects:  it provides pinning 
centers which may couple to the viscous effects even at 
${\bf k=0}$, but it also disrupts the growth in crystalline 
order which yields the viscous effects in the first place. 
 With translational invariance destroyed, the flux-flow 
formula no longer applies.  
 We will examine the effects of disorder at low order in 
perturbation theory within the Kubo formalism.  

 We can in principle take the direct approach at ${\bf k=0}$ 
using the Green's functions with the explicit electric field 
applied.  
 However, it is more complicated than it was in the absence 
of disorder. 
 We confirmed for the pure system that we could use solely 
$n=0$ states from the outset.  
 So long as the electric field was explicitly applied, taking 
the $\tilde \alpha / \hbar \omega_0 \rightarrow 0$ limit in the 
beginning ({\it i.e.} using Green's functions with only $n=0$ 
states) or at the end yielded the same results. 
 However, the same is not true in the presence of disorder.  
 The time integrals here are more involved, and taking the 
$\tilde \alpha / \hbar \omega_0 \rightarrow 0$ limit in the 
beginning and at the end gives different results. 
 With this complication, the direct approach is no better than 
using the Kubo formalism. 

\subsection{Point disorder}
 
In the experiments by Fendrich {\it et al.} \cite{Fendrich} the 
point defects induced by irradiation were seen to introduce an 
additional term to the $ab$ plane conductivity which was not of 
the flux-flow form. 
 They analyzed their resistivity data using the form 
$\rho=(1/\rho_f+1/\rho_P)^{-1}$ where $\rho_f$ is the flux-flow 
resistivity and $\rho_P$ the resistivity due to point defects.  
 The latter had an activated form: 
$\rho_P = \rho_0 ~{\rm exp}\{-U(T,H)/T \}$, and 
the field dependence of the ``plastic energy" was 
$U \sim H^{-0.7 \pm 0.1}$.   

 Let us consider the effect of uncorrelated disorder on the 
conductivity calculated from TDGL. 
 We will model the point defects by adding a quenched random mass 
term to the free-energy functional 
\begin{equation}
F_R[\Psi]= \int d^3{\bf r} \Biggl[
\ldots
+\left(\alpha + \alpha_R({\bf r}) \right) |\Psi|^2 
+ \ldots 
\Biggr],
\label{free-energy-dis}
\end{equation}
where the random variables $\alpha_R({\bf r})$ have zero average  
\begin{equation}
\overline{\alpha_R({\bf r}) } = 0
\end{equation}
and $\delta$-function correlations
\begin{equation}
\overline{ \alpha_R({\bf r})~\alpha_R({\bf r^{\prime}})} = 
{W_0 \over 2} ~\delta({\bf r}-{\bf r^{\prime}}) .
\label{disorder} 
\end{equation}
The associated disordered TDGL can be written as 
\begin{equation}
\Psi({\bf r},t) = \int d {\bf r^{\prime}} \int d t^{\prime} 
G({\bf r},t;{\bf r^{\prime}}, t^{\prime}) \left[ 
\eta({\bf r^{\prime}},t^{\prime}) 
-\alpha_R({\bf r^{\prime}}) \Psi({\bf r^{\prime}},t^{\prime}) 
-\beta |\Psi({\bf r^{\prime}}, 
t^{\prime})|^2\Psi({\bf r^{\prime}}, t^{\prime}) \right]
\label{TDGL-inverted-dis}
\end{equation} 
or symbolically as 
\begin{equation}
\Psi_1 = G_{1,2} \eta_2 
-\alpha_{R2} G_{1,2}\Psi_2
-\beta G_{1,2}\Psi^*_2 \Psi_2 \Psi_2.
\label{symbol-2}
\end{equation}
We have simply added the new term to eq. 
(\ref{TDGL-inverted-2}), hence $G_{1,2}$ remains the same, 
that is, the Green's function of the pure, linearized system. 

 Iterating this equation with $\beta=0$ gives
\begin{equation}
\Psi_1 = G_{1,2} \eta_2 
- \alpha_{R2} G_{1,2}G_{2,3} \eta_3
+\alpha_{R2}\alpha_{R3}G_{1,2} G_{2,3} G_{3,4} \eta_4 
+O(\alpha_R^3).
\label{iterate-a}
\end{equation}
Diagrammatically, we will represent $\alpha_R$ by a small 
square. 
 There will be one arrow (Green's function) pointing into 
it and one out of it. 
 The disorder averaging pairs up the $\alpha_R$'s, we 
represent this feature by a dashed line connecting the 
two squares. 
 Because of the spatial $\delta$ function in the correlation 
of the disorder (\ref{disorder}), squares connected by a 
dashed line represent the same point in space but different 
points in time. 
 Thus as far as spatial integrals are concerned, diagrams in 
this perturbation expansion are identical to those found 
expanding the nonlinear term. 
 It is the time integrals that make it different.  

\begin{figure}
\centerline{
\epsfxsize=7cm \leavevmode \epsfbox{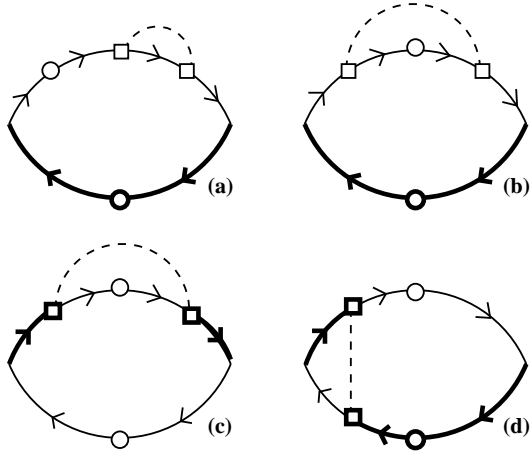}}
\caption{ Some diagrams at first-order in the disorder strength. 
Diagram \ref{order-v}(a) is Green's function renormalizing; 
diagrams \ref{order-v}(b) and (c) are noise renormalizing; and 
diagram \ref{order-v}(d) is vertex-renormalizing.  The last is 
not only ``down" but also zero at ${\bf k=0}$.  }
\label{order-v}
\end{figure}

Figure \ref{order-v} shows some diagrams that arise at 
the first order in an expansion in the disorder strength 
$W_0$. 
 Note that if we were to draw the two points connected by 
the dashed line as one point, then we would end up with 
diagrams very much like those in Figure \ref{order-b}.  
 (They would still differ by the number of circles.) 
 In fact, the diagram in Fig. \ref{order-v}(d) gives zero 
contribution at ${\bf k=0}$ for both $\sigma_{xx}$ and 
$\sigma_{zz}$ for the same reasons as that in Fig. 
\ref{order-b}(b). 
 The diagram in Fig. \ref{order-v}(a) is Green's function 
renormalizing, those in (b) and (c) noise renormalizing, and 
that in (d) vertex-renormalizing.  

Certain basic facts about the calculations do not change in 
the presence of disorder.  
 For instance, in the calculation of 
$\sigma_{xx}(k_x,k_y=0,k_z)$, it remains true 
that there is a maximum of two $n\geq1 $ Green's functions in 
diagrams contributing to the $\tilde \alpha /\hbar \omega_0 
\rightarrow 0$ limit. 
 Hence the diagram in Fig. \ref{order-v}(d) does not contribute 
to this order.  
 Another invariant is that either the two $n \geq 1$ Green's 
functions meet each other at a noise average (as in Fig. 
\ref{order-v}(a) and (b)) or they each meet a vertex (as in 
Fig. \ref{order-v}(c)). 
 The difference is now there are two types of vertices, the 
original quartic vertices and the new disorder induced vertices. 
 From the case in which the $n\geq 1$ Green's functions meet 
at a noise average, one gets the disordered analog of the 
flux-flow formula with $\langle |\Psi_0|^2 \rangle$ replaced 
by its disordered counterpart. 
 (This particular subset of diagrams yields the $1/\rho_f$ in 
the Fendrich {\it et al.} \cite{Fendrich} analysis.)  
 From the case in which the two Green's functions meet quartic 
vertices, one gets the disordered analog of eq. 
(\ref{not-flux-flow}).  

 Then there are new cases, like the noise-renormalizing diagram 
shown in Fig. \ref{order-v}(c). 
Evaluating it yields 
\begin{equation}
\sigma_{xx}^{7(c)}(k_z) = \sigma_{xx}^{(G)}({\bf 0}) ~{(\pi -2) 
z_0 \over 2 } ~ \left[1 +{1 \over 2} k_z^2 \ell_c^2 \right]^{-2}. 
\end{equation}
Note that we have factored out the pure Gaussian result; it 
multiplies a numerical factor and $z_0$, the dimensionless 
expansion parameter \cite{Fujita} for point disorder
\begin{equation}
z_0= { W_0 \over 16 \pi \ell^2 \xi_c {\tilde \alpha}^2}.
\end{equation}
We have also included the diagram's $k_z$ dependence. 
 The $c$-axis length scale associated with this diagram, as 
well as those in Fig. \ref{order-v} (a) and (b), is $\ell_c$.    
  As in the pure case, we can send the external momentum $k_z$ 
through channel made entirely of $n=1$ states.  
 Recall that the points connected by the dashed line are the 
same point in space; consequently, in Fig. \ref{order-v}(c) the 
dashed line acts as a ``short" allowing the external momentum 
$k_z$  to pass through a channel comprised solely of $n \geq 1$ 
states. 

By once again factoring out the higher Landau level Green's 
functions, this last calculation might be generalized to      
\begin{eqnarray}
\Sigma^{(1)}_{xx}(k_x,k_z) &=& {e^{*2} W_0 \over 16 k_B T 
m_{ab}^2}\int d({\bf r}-{\bf r^{\prime}}) \int d(t-t^{\prime})
 \int d {\bf r_5} \int dt_5 \int dt_6
~{\rm e}^{i k_x (x- x^{\prime})+ i k_z (z- z^{\prime}) } 
\nonumber \\ 
&&\Bigl(P_{1x}+P_{2x}^*\Bigr)\Bigl(P_{3x}+P_{4x}^*\Bigr)  
\Biggl\{ G^*({\bf r_2},t;{\bf r_5},t_5)G({\bf r_3},
t^{\prime};{\bf r_5},t_6) 
\nonumber \\
&& \langle \Psi_0 ({\bf r_1},t) 
\Psi_0^*({\bf r_5},t_5)  \Psi_0({\bf r_5},t_6) 
\Psi_0^*({\bf r_4},t^{\prime}) 
\rangle_c  + c.c. \Biggr\} ,
\label{four-point}
\end{eqnarray}      
where we imagine the remaining LLL object as being fully 
renormalized. 
 The disorder average was performed with the result that 
the order parameter fields $\Psi_0({\bf r_5},t_5)$ and 
$\Psi_0^*({\bf r_5},t_5)$ are at the same point in space but 
different points in time.  
 Note that the LLL object here is a four-point quantity and 
thus probes viscous effects even at ${\bf k=0}$.      
 The resummation suggested in eq. (\ref{four-point}) is just 
one example of a new feature induced by the defects.   
     
\subsection{Correlated Disorder}

We model columnar defects lying parallel to the $c$ axis by 
changing the correlation of the disorder from a $3D$ delta 
function to a $2D$ delta function
\begin{equation}
\overline{ a({\bf r})~a({\bf r^{\prime}})
} = {W_1 \over 2} ~\delta(x- x^{\prime})~\delta(y- y^{\prime}) .
\label{cor-dis} 
\end{equation}
For sake of comparison we calculated the diagram in Fig. 
\ref{order-v}(c) for columnar defects, obtaining 
\begin{equation}
\sigma_{xx}^{7(c)}(k_z,0) = \sigma_{xx}^{(G)}({\bf 0}) ~ 
{3 z_1 \over 8 } ~ \left[{8(-2 + (1+k_z^2 \xi_c^2/4)^{1/2}
+  (1+k_z^2 \xi_c^2/4) \over
 3 k_z^2\xi_c^2 (1+k_z^2 \xi_c^2/4)^2}  \right],  
\end{equation}
where we have factored the result as above and $z_1$ is 
the dimensionless expansion parameter for columnar defects
\begin{equation}
z_1= { W_1 \over 16 \pi \ell^2  {\tilde \alpha}^2}. 
\end{equation}
Note that $z_1$ differs from $z_0$ by the absence of $\xi_c$ 
in the denominator of $z_1$, making $z_1$ the larger of the 
two in the $\tilde \alpha \rightarrow 0$ limit.  
 The $c$-axis length scale associated with the diagram 
in Fig. \ref{order-v}(c) is $\xi_c$.    
 Points connected by a dashed line are no longer at the 
same point in space. 
 They have different values of $z$; therefore, the dashed 
line no longer serves as a ``short" for external momentum 
$k_z$ and we can no longer send  $k_z$ through an 
exclusively $n \geq 1$ channel. 
 The switch from the $\ell_c$ length scale for point 
defects to $\xi_c$ for columnar defects in the evaluation of 
theses diagrams is in accord with the suggestion that correlated 
defects align pancake vortices enhancing their integrity.

We model planar defects by changing the disorder correlations 
to 
\begin{eqnarray}
\overline{ a({\bf r})~a({\bf r^{\prime}})} 
&=& {W_{2x} \over 2} ~\delta(x- x^{\prime}); \nonumber \\
\overline{ a({\bf r})~a({\bf r^{\prime}})} 
&=& {W_{2y} \over 2} ~\delta(y- y^{\prime}) ,
\label{dis-plane} 
\end{eqnarray}
where the former models planar defects parallel to the $yz$ 
plane and the latter those parallel to the $xz$ plane. 
 When the electric field is in the $x$ direction, the vortices 
move in the $y$ direction.  
 One might expect different results for vortices moving 
parallel to or perpendicular to the defects.
 Evaluation of the diagram in Fig. \ref{order-v}(c) is then  
\begin{eqnarray}
\sigma_{xx}^{7(c)}(k_z,0) &=& \sigma_{xx}^{(G)}({\bf 0}) ~ 
{3 \sqrt{\pi} z_{2x} \over 8 \sqrt{2} } ~ 
\left[{8(-2 + (1+k_z^2 \xi_c^2/4)^{1/2}+  
(1+k_z^2 \xi_c^2/4) \over
 3 k_z^2\xi_c^2 (1+k_z^2 \xi_c^2/4)^2}  \right];
\nonumber \\
\sigma_{xx}^{7(c)}(k_z,0) &=& \sigma_{xx}^{(G)}({\bf 0}) ~ 
{3 \sqrt{\pi} z_{2y} \over 8 \sqrt{2} } ~ 
\left[{8(-2 + (1+k_z^2 \xi_c^2/4)^{1/2}+  
(1+k_z^2 \xi_c^2/4) \over
 3 k_z^2\xi_c^2 (1+k_z^2 \xi_c^2/4)^2}  \right], 
\end{eqnarray}
where 
\begin{equation}
z_{2i} = { W_{2i} \over 16 \pi \ell  {\tilde \alpha}^2}
\end{equation}
(with $i=x$ or $y$) are the dimensionless expansion parameters 
for planar defects.
 For this particular diagram we find no difference between 
the $xz$ and $yz$  planar defects.   
 The planar defect expansion parameter $z_{2i}$ has the same 
$\tilde \alpha$ dependence and so the same temperature 
dependence as $z_1$.  
 On the other hand, $z_1$ has an additional $\ell$ in the 
denominator so the magnetic field dependence of these 
expansion parameters is different.   
 
\section{Summary}

We have investigated the conductivity in the vortex-liquid 
regime via perturbation theory. 
 Using an approach in which an electric field is applied 
explicitly we showed that within the lowest Landau level 
(LLL) approximation the uniform $ab$ plane conductivity, 
$\sigma_{xx}({\bf 0})$, is proportional to $\langle | 
\Psi_0|^2 \rangle$ with a temperature-independent 
proportionality constant, to all orders in perturbation theory.  
 This result  elevates the derivation of the flux-flow formula 
to non-zero temperatures. 
 We identified the subset of diagrams that yields the same 
result calculated within the Kubo formalism and conjectured 
that the remaining diagrams must cancel.  
 We verified this cancellation to second order in perturbation 
theory and also demonstrated that the cancellation does not 
extend to the nonlocal conductivity $\sigma_{xx}({\bf k})$, 
allowing viscous effects to enter. 

 When it comes to the nonlocal conductivity and the issue 
of characteristic lengths, the present work has certainly 
raised more questions than it has answered. 
 For instance, the lowest-order (Gaussian) diagram of the 
transverse $ab$-plane conductivity already revealed a long, 
temperature-dependent length; while all of the diagrams for 
the longitudinal $ab$-plane conductivity individually had 
short, temperature-independent length scales.  
 If the longitudinal length scale is to grow long, as is 
suggested by its identification with shearing effects of a 
growing crystal, it will be through the combined effect of 
many diagrams.  
 The nature of the diagrams that may lead to such an effect 
have been identified, but what remains unclear is whether 
the longitudinal and transverse length scales are independent 
or related.  

There is a similar question regarding the $c$-axis. 
 Every diagram contributing to the LLL approximation of 
$\sigma_{zz}$ has $\xi_c$, the temperature-dependent 
correlation length, as its length scale.  
 On the other hand, at each order in the perturbative 
expansion of $\sigma_{xx}$, the $c$-axis length scale is 
$\ell_c$, a temperature-independent magnetic length. 
 If $\sigma_{xx}(z-z^{\prime})$ is to become substantially 
nonlocal as $T$ is lowered, it will be through the combined 
effect of various orders. 
 If the two length scales are long, it is unclear whether 
they will turn out to be essentially the same or distinct.  
 In addition, there is the question of whether or not the 
$ab$-plane and $c$-axis length scales grow independently. 
 Finally, these matters must be readdressed with regard 
to the nonlocal resistivity which may have its own 
distinct length scales. 

We also examined the effect of disorder on the $ab$-plane 
conductivity. 
 While there was a subset of diagrams that lead to the 
disordered analog of the flux-flow formula, there were 
other contributions that coupled to the viscous effects 
even at ${\bf k=0}$. 
 We also showed that diagrams contributing to the LLL 
approximation had as their $c$-axis length scale $\ell_c$ 
in the presence of point defects and $\xi_c$ in the presence 
of columnar and planar defects. 

 Missing from this work is any consideration of what 
vector-potential fluctuations might do.  
 Including them would require a new kind or vertex, 
one involving a $\Psi$, a $\Psi^*$ and an ${\cal A}_{\mu}$, 
where ${\cal A}_{\mu}$ is the fluctuating part of the 
vector potential. 
 The arguments leading to the flux-flow result may no 
longer hold when these new vertices are added.  
 Also lacking are the effects of Maxwell's equations 
which are needed to form a complete set of equations. 
 At low order these lead to back-flow effects and at 
higher orders may produce lattice effects.  
 There are many challenging aspects of this problem 
to be resolved.    

\section{Acknowledgements}

We would like to thank the authors of Ref. \cite{Lopez} 
for fruitful correspondence.
 We would also like to thank M.J.W. Dodgson, D.A. Huse, 
H. Jensen, A.J. McKane, N.K. Wilkin and J. Yeo for useful 
discussions. 
 TB acknowledges the support of the Engineering and Physical 
Science Research Council under grant GR/K53208 and of the 
National Science Foundation under grant DMR9312476.


\bibliographystyle{unsrt}

\end{document}